# Leveraging Digital Twin and Machine Learning Techniques for Anomaly Detection in Power Electronics Dominated Grid

Ildar Idrisov
*Center for Energy Science and Technology*
*Skolkovo Institute of Science and Technology*
Moscow, Russia
i.idrisov@skoltech.ru

Divine Okeke
*Center for Energy Science and Technology*
*Skolkovo Institute of Science and Technology*
Moscow, Russia
Divine.Okeke@skoltech.ru

Abdullatif Albaseer
*Division of Information and Computing Technology*
*College of Science and Engineering*
*Hamad Bin Khalifa University*
Doha, Qatar
aalbaseer@hbku.edu.qa

Mohamed Abdallah
*Division of Information and Computing Technology*
*College of Science and Engineering*
*Hamad Bin Khalifa University*
Doha, Qatar
moabdallah@hbku.edu.qa

Federico M. Ibanez
*Center for Energy Science and Technology*
*Skolkovo Institute of Science and Technology*
Moscow, Russia
FM.Ibanez@skoltech.ru

*Abstract*— Modern power grids are transitioning towards power electronics-dominated grids (PEDG) due to the increasing integration of renewable energy sources and energy storage systems. This shift introduces complexities in grid operation and increases vulnerability to cyberattacks. This research explores the application of digital twin (DT) technology and machine learning (ML) techniques for anomaly detection in PEDGs. A DT can accurately track and simulate the behavior of the physical grid in real-time, providing a platform for monitoring and analyzing grid operations, with extended amount of data about dynamic power flow along the whole power system. By integrating ML algorithms, the DT can learn normal grid behavior and effectively identify anomalies that deviate from established patterns, enabling early detection of potential cyberattacks or system faults. This approach offers a comprehensive and proactive strategy for enhancing cybersecurity and ensuring the stability and reliability of PEDGs.

## I. INTRODUCTION

The widespread integration of renewable energy sources, electric transport [1] and energy storage systems (ESS) [2] into power grids led to a new energy paradigm, where traditional distribution systems are heavily relying on power electronics devices, often referred to as power electronics-dominated grids (PEDGs) [3], [4]. This shift introduces greater complexity and elevates the importance of both device- and system-level control strategies to ensure grid resilience, reliability, and operational stability [5].

To deal with this complexity wide-area measurement systems and Internet of Things technologies have been introduced. Such technologies allow to realize effective monitoring and control of PEDGs, but conversely amplify the degree of the cyber system complexity due to the use of multiple industrial controllers, communication protocols, intelligent electronics devices, smart meters, and phasor measurement units. This evolution effectively transforms the modern electric distribution system into a complex and critical energy cyber-physical system [6], where the physical grid infrastructure is deeply linked with its digital control and communication systems [7]. These new infrastructure lead to the vulnerability of cyberattacks which put under cybersecurity risks, where even a small false injected data as an attack can be resulted in power outage [8], [9].

The cyber-physical system security field primarily focuses on two key approaches for mitigating cyberattacks: attack detection and mitigation, as well as implementation of resilient control systems. Among these approaches, there are i) signal-based attack detection methods, ii) model-based detection approaches, and iii) data-driven algorithms [10].

The first method, Signal-based attack detection, which is relatively straightforward and frequently used in microgrids, depends on real-time observation of signals from communication links [11], [12], [13]. However, a major drawback of this approach is its inability to sufficiently assess the correlation between recorded data and control signals, an essential element for dependable attack detection [14].

Alternatively, ii) model-based detection methods utilize mathematical models of the system to spot deviations and abnormal activities that could indicate cyber-attack. These methods are designed to compare expected measurements with actual ones, and have proven effective against various kinds of cyber-attacks [15], [16]. However, identifying sophisticated attacks and developing precise models can be complicated and computationally expensive, which needs a thorough understanding of typical system behavior patterns. Additionally, this method does not account for changes or deviations in physical parameters and model characteristics, such as degradation effects, thus affect long-term prediction accuracy.

A third approach, data-driven detection, utilizes machine learning or statistical techniques to infer system models from historical data and measurement signals, thereby enabling the identification of malicious behavior [17], [18]. A key benefit of this technique is their adaptability to diverse network environments and their ability to uncover hidden patterns without predefined rules. However, these approaches can be computationally intensive, especially with large datasets, requiring substantial processing resources.

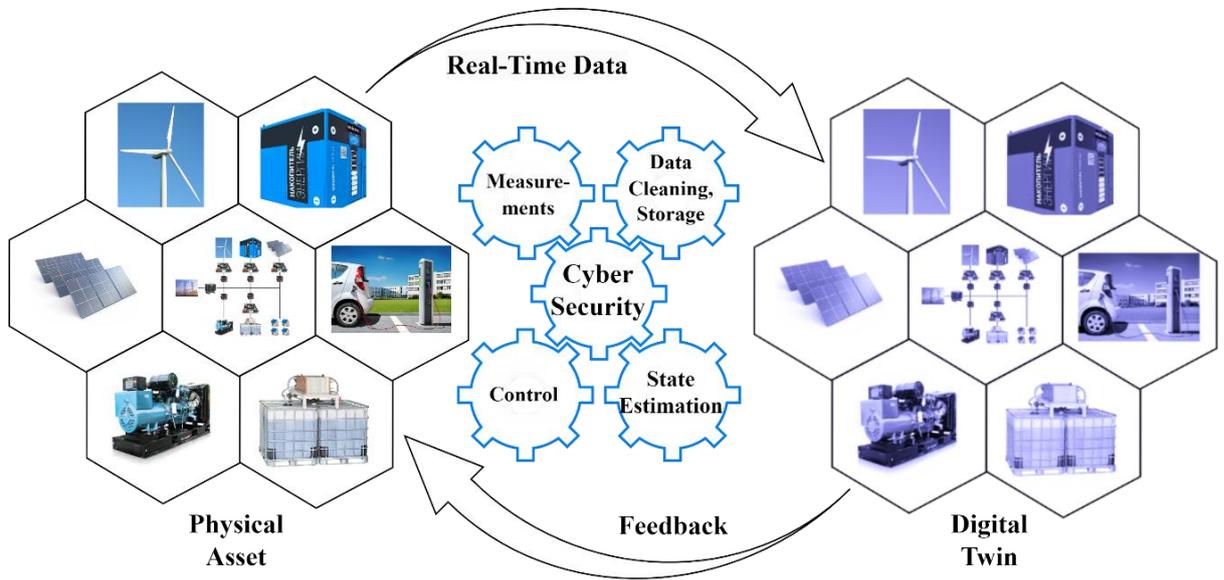

Fig. 1 Simplified illustration of digital twin definition.

DT technology in this context offers the potential to revolutionize cyberattack detection and mitigation in PEDGs. First developed within the aerospace and aviation industries, DT serves as a sophisticated digital representation of a physical asset or process, maintaining a continuous exchange of information between them to deliver actionable insights. Its transformative capabilities have paved the way for significant advancements in various domains, including the energy sector [19]. As depicted in Fig. 1, a digital twin establishes a seamless connection between the physical components and processes of the system and their virtual replica, enabling real-time, bidirectional data flow to enhance monitoring, analysis, and optimization.

In this work, we combine the power of DT technology with data-driven methods. The DT provides an enhanced model, which is accurately mirror the current state of the power system, while robust data-driven models capable of detecting anomalies in power system measurements, delivered by DT. We are using both traditional machine learning and deep learning approaches by addressing challenges such as class imbalance and leveraging cross-validation techniques.

II. DIGITAL TWIN OF POWER ELECTRONICS DOMINATED GRID

PEDGs exhibit faster, more complex dynamics compared to traditional systems, requiring advanced tools for monitoring, control, and cybersecurity. By integrating dynamic state estimation (DSE) techniques, DT becomes essential for capturing fast-changing states such as voltages, currents, and converter dynamics [E]. Accurate DSE enables the DT to reflect the real-time behavior of the grid with high fidelity, making it possible to detect anomalies, such as sudden voltage sags, harmonic distortions, or frequency instability, that are often precursors to cyberattacks or system faults.

To build DSE enabled digital twin, a suitable distributed generations dynamic models should be included along with transmission lines and power inverters. Among them photovoltaics [20], wind turbines [21] and ESS [22] which is widely discussed in the literature and is not within the scope of this work.

A. *Inverter DT Model Formulation*

The inverter generation interface is a complex dynamic system, where the dominant income comes from phase-locked-loop (PLL), current control loop and LCL output filter [4], [23].

In this work, 4-quadrant inverter in grid following mode applied. In order not to be computational expensive, real-time aggregated model have been realized according the diagram in Fig. 2. The inverter behaves as current source and deliver required active and reactive current into the grid.

The current control loop is implemented using a $D-Q$ decoupling scheme reliant on PI regulators, with parameters specifically selected to offset the dynamics of the output LCL filter [23]. All parameters of the current control loop, PLL, and output filter should be derived from the actual device installed in the grid to precisely replicate all processes. Functioning in grid-following mode, current control loop receives references for active and reactive currents $i_d^*, i_q^*$ from intelligent electronic devices (IED) and deliver power to the point of connection (POC):

$$p = i_d v_{poc} \cos(\rho) + i_q v_{poc} \sin(\rho),$$
$$q = -i_q v_{poc} \cos(\rho) + i_d v_{poc} \sin(\rho). \quad (1)$$

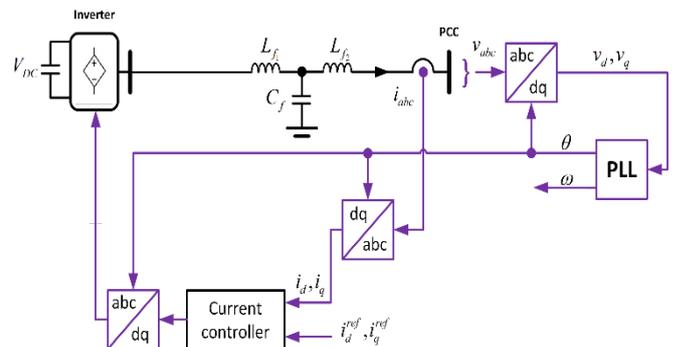

Fig. 2 Model of inverter in grid-following mode.

## B. Load DT Model Formulation

The typical consumer in distribution grid generally can be modelled as a function of positive-sequence voltage:

$$P_l = P_0(V/V_0)^{n_p},$$
$$Q_l = Q_0(V/V_0)^{n_q}, \quad (2)$$

where $n_p$ and $n_q$ are exponents (usually between 1 and 3) controlling the nature of the load. $V_0$ is the initial positive sequence voltage. $P_0$ and $Q_0$ are the initial active and reactive powers at the initial voltage $V_0$. The consuming active $P$ and reactive $Q$ power follows the references received from IED at POC.

## C. Transmission Lines

Power transmission lines (TL) in distribution grid have a relatively short length and low rated voltages, conduction currents are small compared to load currents. Therefore, in typical electrical calculations of these networks, the capacitive line conductivities may not be taken into account [24]. Accordingly, the simplified power line replacement scheme in the model will take the form shown in Fig. 3.

## III. MACHINE LEARNING METHODOLOGY

In this section, we detail the methodologies employed in our paper, including data acquisition, preprocessing, feature selection, model development, and evaluation techniques. To improve the functionality and safety of DT based state estimator robust models developed, capable of detecting anomalies in power system measurements using both traditional machine learning and deep learning approaches by addressing challenges such as class imbalance and leveraging cross-validation techniques.

### A. Data Preprocessing

Data preprocessing is essential to ensure the integrity and suitability of the data for modeling. The following steps were conducted:

*1) Data Cleaning:* The raw data contained inconsistencies due to variations in data formats and potential errors during data collection. To address these issues, we first ensured accurate parsing of data by specifying the correct delimiter when importing the datasets. We standardized numerical values by replacing locale-specific decimal separators (e.g., commas) with periods. We then converted all numerical columns to appropriate data types to facilitate numerical computations. We identified and removed rows containing missing or invalid data entries to maintain data quality.

*2) Data Labeling:* To prepare the data for supervised learning algorithms, we assigned binary labels to each data instance, with 0 indicating normal operation and 1 indicating an attack or anomaly. Then, we merged the labeled normal and attack datasets into a single cohesive dataset for subsequent analysis. To ensure that all features contribute equally to the model training process, feature scaling was applied as follows. We employed z-score normalization to rescale features to have a mean of zero and a standard deviation of one, defined as:

$$z = \frac{x - \mu}{\sigma}, \quad (3)$$

where $x$ is the original feature value, μ is the mean of the feature, and $\sigma$ is the standard deviation.

Exploratory data analysis (EDA) was conducted to gain insights into the characteristics of the data. We calculated descriptive statistics, including mean, median, variance, and interquartile ranges for each feature. We created time-series plots of power outputs and voltage levels to observe trends, patterns, and potential anomalies. Accordingly, we computed the Pearson correlation coefficients between features to identify multicollinearity, which could adversely affect certain modeling techniques.

We performed a feature selection to enhance model performance and interoperability by utilizing a Random Forest classifier to estimate feature importance based on the Gini impurity decrease criterion. The features have been ranked according to their importance scores to identify the most significant predictors influencing the target variable. We considered excluding features with low importance scores to reduce dimensionality and computational complexity without compromising model accuracy.

Given the potential imbalance between normal and attack instances in the dataset, we implemented techniques to address this issue. First, we augmented the dataset by introducing controlled Gaussian noise to the existing data points, thereby increasing variability and aiding the model in generalizing better. We applied Synthetic Minority Over-sampling Technique (SMOTE) to generate synthetic examples of the minority class (attack data), balancing the class distribution and mitigating the risk of model bias towards the majority class. We randomized the order of data instances to eliminate any inherent order that could bias the model. We divided the dataset into training and testing sets using stratified sampling to preserve the original class proportions in both subsets.

### B. Traditional Machine Learning Approach

*1) Random Forest:* First, a Random Forest [25] classifier was employed as a traditional machine learning model for anomaly detection. We set parameters such as the number of trees (estimators), maximum tree depth, minimum samples required to split an internal node, and minimum samples required at a leaf node. We trained the model on the preprocessed and scaled training data, allowing it to learn patterns and relationships between features and the target variable.

*2) Long Short-Term Memory (LSTM):* An LSTM [26] neural network was utilized to capture temporal dependencies inherent in time-series data. We reshaped the dataset into sequences appropriate for LSTM input, with each sequence representing a time window of observations. The sequences and corresponding labels were converted into tensors compatible with the deep learning framework used for model implementation. We configured to accept input sequences of feature vectors. We included one or more LSTM layers to model sequential dependencies and retain information over time steps. A dense layer with appropriate activation (e.g., sigmoid for binary classification) was added to produce the final output.

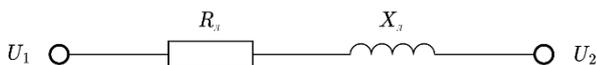

Fig. 3 Simplified scheme of TL block of distribution grid model.

We employed binary cross-entropy loss suitable for binary classification tasks. The Adam optimization algorithm was used for efficient gradient descent. We adjusted hyperparameters such as learning rate, number of epochs, batch size, and the number of hidden units to optimize model performance.

We assessed the model's performance on a validation set to monitor training progress and prevent overfitting. The final model was evaluated on the test set to obtain an unbiased estimate of its generalization performance. We applied *k*-fold cross-validation to ensure the robustness of the results and to mitigate the effects of any data partitioning bias where we first divided the dataset into *k* equally sized folds; iteratively trained the model on *k* - 1 folds and validated it on the remaining fold. We ensured that each fold maintained the same class distribution as the entire dataset to reflect the model's performance across different subsets accurately. The average and standard deviation of performance metrics are averaged across all folds to provide a comprehensive assessment.

## IV. ANOMALY DETECTION VALIDATION

To validate threat detection based on collaborative work of DT and ML methods, the hardware-in-the-loop technique is utilized with Man-in-the-middle (MITM) attack. The real microgrid structure utilized as detailed electromagnetic transient model within a real-time simulator (RTDS Novacor) representing consumers, solar generation, wind generation, and ESS with the appropriate power electronics interface. To reproduce real power flow, logged generation and consuming profiles are used in time as referenced values for power inverters and loads.

DT represented by averaged real-time model implemented in separate real-time machine OPAL-RT 5600. To accurate reflect the state of real grid, DT leverages sinusoidal analog voltages at the point of common coupling in the head of the feeder along with digital measurements transmitted from IED by IEC 60870-5-104 protocol. The measured power values received act as reference points for the loads and power inverters in the DT. This configuration guarantees that the overall DT model accurately represents the real-time condition of the physical asset simulated by the RTDS simulator. Fig. 4 depicts this detailed test bench configuration.

### A. Dataset Description

The datasets utilized in this study comprise time-series measurements collected from a power system and DT under both normal operating conditions and simulated attack scenarios. The data include various electrical parameters, such as:

- **Power Outputs**: Photovoltaic active and reactive power ($P_{pv}$, $Q_{pv}$), battery active and reactive power ($P_{batt}$, $Q_{batt}$), wind active and reactive power ($P_w$, $Q_w$) and consumer active and reactive powers ($P_n$, $Q_n$) – all received from IED.
- **Voltage Levels**: Voltage virtual measurements estimation from DT at different nodes within the whole feeder ($V_1$ to $V_6$).
- **Frequency**: Frequency virtual measurements estimation from DT at different nodes within the whole feeder ($F_1$ to $F_6$).
- **Time Stamps**: Temporal markers indicating the time of each measurement.

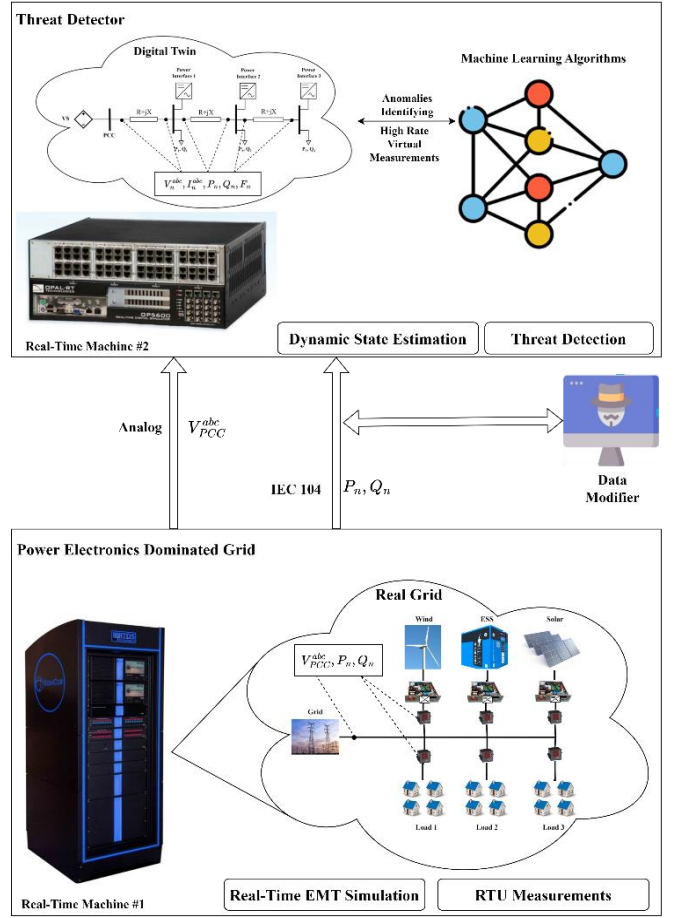

Fig. 4 Test bench system for anomalies detection.

### B. Machine Learning Setup

All experiments were conducted under the following conditions. For Random Forest Classifier we set the number of estimators to 100, the maximum depth to 10, the minimum samples split to 10, and the minimum samples leaf to 5. For LSTM Neural Network, we used 32 and 64 hidden Units. The learning rate is adjusted between 0.0001 and 0.001 with a number of epochs between 50 and 150 based on convergence observations. The batch size is selected to optimize training efficiency and convergence stability. A k-fold cross-validation was performed with $k = 10$ to estimate the model's generalization capability and to validate its stability across different subsets of data. We applied methods such as dropout and weight decay to prevent overfitting in the neural network. We implemented early stopping based on validation loss to prevent overfitting by halting training when no improvement was observed.

### C. Performance Metrics

To evaluate and compare the models, several performances metrics were employed:

*1) Accuracy:* The proportion of correctly classified instances among the total instances evaluated:

$$Accuracy = \frac{TP + TN}{TP + TN + FP + FN}, \quad (4)$$

where TP, TN, FP, and FN represent true positives, true negatives, false positives, and false negatives, respectively.

*2) Precision: The proportion of correctly classified instances among the total instances:* The ratio of true positives to the sum of true positives and false positives:

$$Precision = \frac{TP}{TP + FP}. \quad (5)$$

*3) Recall:* The ratio of true positives to the sum of true positives and false negatives:

$$Recall = \frac{TP}{TP + FN}. \quad (6)$$

*4) F1-Score:* The harmonic mean of precision and recall:

$$F1 - Score = 2 \times \frac{Precision \times Recall}{Precision + Recall}. \quad (7)$$

*5) Confusion Matrix:* A matrix that provides a detailed breakdown of correct and incorrect classifications, allowing for the analysis of the types of errors made by the model.

### D. Threat Detection Results

The performance of both the Random Forest and LSTM models was evaluated under two scenarios: using only data from measurement devices and utilizing the DT enhanced data. The results of these evaluations are presented in Table TABLE I.

TABLE I. ML PERFOMANCE METRICS

| Method | Performance Metrics Without DT | | | |
|---|---|---|---|---|
| | *Accuracy* | *Precision* | *Recall* | *F1-Score* |
| Random Forest | 0.7434 | 0.73 | 0.77 | 0.75 |
| LSTM | 0.8688 | 0.8325 | 0.8183 | 0.8254 |
| Method | Performance Metrics With DT | | | |
| Random Forest | 0.8692 | 0.7380 | 0.8748 | 0.8006 |
| LSTM | 0.9159 | 0.9417 | 0.8669 | 0.9028 |

The evaluation of threat detection performance revealed that integrating DT-enhanced data significantly improved the accuracy and overall metrics for both Random Forest and LSTM models.

## V. CONCLUSIONS

This research successfully demonstrates the synergistic potential of digital twin technology and machine learning for enhanced anomaly detection in power electronics-dominated grids . The DT provides a real-time, data-rich environment mirroring the physical system, enabling ML algorithms, specifically Random Forests and LSTMs, to learn normal operating patterns and identify subtle deviations indicative of cyberattacks or system faults.

Additionally, DT technology allows for training ML models with limited statistical data, leveraging its ability to simulate and generate robust datasets. DT not only a tool for performance enhancement but also a strategic enabler for effective threat detection in data-constrained environments. Future work should focus on optimizing model performance for higher grid dimension to fully realize the potential of this method.

ACKNOWLEDGMENT

This work was supported by HBKU-Skoltech project: Smart detection of cyber-attacks in electric metering using validated machine learning algorithms.

REFERENCES

[1] Y. Khan, I. Idrisov, M. Pugach, and F. Martin Ibanez, "Real-Time Analysis of Battery State of Health in Supercapacitor-Battery Hybrid Systems for Electric Motorcycles," IEEE Access, vol. 12, pp. 151403–151414, 2024, doi: 10.1109/ACCESS.2024.3478375.

[2] F. D. Hernandez, F. Ibanez, R. Samanbakhsh, and R. Velazquez, "A Comparative Study of Energy Storage Systems based on Modular Multilevel Converters," in IECON 2021 – 47th Annual Conference of the IEEE Industrial Electronics Society, Toronto, ON, Canada: IEEE, Oct. 2021, pp. 1–5. doi: 10.1109/IECON48115.2021.9589539.

[3] A. Ipakchi and F. Albuyeh, "Grid of the future," IEEE Power Energy Mag., vol. 7, no. 2, pp. 52–62, Mar. 2009, doi: 10.1109/MPE.2008.931384.

[4] A. A. Nazeri, P. Zacharias, F. M. Ibanez, and I. Idrisov, "Paralleled Modified Droop-Based Voltage Source Inverter for 100% Inverter-Based Microgrids," in 2021 IEEE Industry Applications Society Annual Meeting (IAS), Oct. 2021, pp. 1–8. doi: 10.1109/IAS48185.2021.9677128.

[5] A. Khan, M. Hosseinzadehtaher, M. B. Shadmand, S. Bayhan, and H. Abu-Rub, "On the Stability of the Power Electronics-Dominated Grid: A New Energy Paradigm," IEEE Ind. Electron. Mag., vol. 14, no. 4, pp. 65–78, Dec. 2020, doi: 10.1109/MIE.2020.3002523.

[6] Y. Ding, B. Wang, Y. Wang, K. Zhang, and H. Wang, "Secure Metering Data Aggregation With Batch Verification in Industrial Smart Grid," IEEE Trans. Ind. Inform., vol. 16, no. 10, pp. 6607–6616, Oct. 2020, doi: 10.1109/TII.2020.2965578.

[7] A. Saad, T. Youssef, A. T. Elsayed, A. Amin, O. H. Abdalla, and O. Mohammed, "Data-Centric Hierarchical Distributed Model Predictive Control for Smart Grid Energy Management," IEEE Trans. Ind. Inform., vol. 15, no. 7, pp. 4086–4098, Jul. 2019, doi: 10.1109/TII.2018.2883911.

[8] G. Liang, S. R. Weller, J. Zhao, F. Luo, and Z. Y. Dong, "The 2015 Ukraine Blackout: Implications for False Data Injection Attacks," IEEE Trans. Power Syst., vol. 32, no. 4, pp. 3317–3318, Jul. 2017, doi: 10.1109/TPWRS.2016.2631891.

[9] Su Sheng, Wang Yingkun, Long Yuyi, Li Yong, and Jiang Yu, "Cyber attack impact on power system blackout," in IET Conference on Reliability of Transmission and Distribution Networks (RTDN 2011), London, UK: IET, 2011, pp. 3B3-3B3. doi: 10.1049/cp.2011.0520.

[7] Shafei, H., Li, L., Aguilera, R.P. (2023). A Comprehensive Review on Cyber-Attack Detection and Control of Microgrid Systems. In: Haes Alhelou, H., Hatziargyriou, N., Dong, Z.Y. (eds) Power Systems Cybersecurity. Power Systems. Springer, Cham. https://doi.org/10.1007/978-3-031-20360-2_1

[11] C.-H. Lo and N. Ansari, "CONSUMER: A Novel Hybrid Intrusion Detection System for Distribution Networks in Smart Grid," IEEE Trans. Emerg. Top. Comput., vol. 1, no. 1, pp. 33–44, Jun. 2013, doi: 10.1109/TETC.2013.2274043.

[12] E. Drayer and T. Routtenberg, "Detection of False Data Injection Attacks in Smart Grids Based on Graph Signal Processing," IEEE Syst. J., vol. 14, no. 2, pp. 1886–1896, Jun. 2020, doi: 10.1109/JSYST.2019.2927469.

[13] M. AlSadat and F. M. Ibanez, "Imitatsionnoe modelirovanie sistem elektrosnabzheniya v programme MATLAB," in 2021 IEEE Power & Energy Society General Meeting (PESGM), Jul. 2021, pp. 1–5. doi: 10.1109/PESGM46819.2021.9638164.

[14] S. Tan, P. Xie, J. M. Guerrero, and J. C. Vasquez, "False Data Injection Cyber-Attacks Detection for Multiple DC Microgrid Clusters," Appl. Energy, vol. 310, p. 118425, Mar. 2022, doi: 10.1016/j.apenergy.2021.118425.

[15] S. Soltani, M. Kordestani, P. K. Aghaee, and M. Saif, "Improved Estimation for Well-Logging Problems Based on Fusion of Four Types of Kalman Filters," IEEE Trans. Geosci. Remote Sens., vol. 56, no. 2, pp. 647–654, Feb. 2018, doi: 10.1109/TGRS.2017.2752460.

[16] B. Li, T. Ding, C. Huang, J. Zhao, Y. Yang, and Y. Chen, "Detecting False Data Injection Attacks Against Power System State Estimation With Fast Go-Decomposition Approach," IEEE Trans. Ind. Inform., vol. 15, no. 5, pp. 2892–2904, May 2019, doi: 10.1109/TII.2018.2875529.

[17] M. Ozay, I. Esnaola, F. T. Yarman Vural, S. R. Kulkarni, and H. V. Poor, "Machine Learning Methods for Attack Detection in the Smart

Grid," IEEE Trans. Neural Netw. Learn. Syst., vol. 27, no. 8, pp. 1773–1786, Aug. 2016, doi: 10.1109/TNNLS.2015.2404803.

[18] M. R. Habibi, H. R. Baghaee, T. Dragicevic, and F. Blaabjerg, "Detection of False Data Injection Cyber-Attacks in DC Microgrids Based on Recurrent Neural Networks," IEEE J. Emerg. Sel. Top. Power Electron., vol. 9, no. 5, pp. 5294–5310, Oct. 2021, doi: 10.1109/JESTPE.2020.2968243.

[19] N. Bazmohammadi, A. Madary, J. C. Vasquez, B. Khan, and J. M. Guerrero, "Microgrid Digital Twins: Concepts, Applications, and Future Trends," vol. 10, 2022.

[20] D. O. Machado et al., "Digital twin of a Fresnel solar collector for solar cooling," Appl. Energy, vol. 339, p. 120944, Jun. 2023, doi: 10.1016/j.apenergy.2023.120944.

[21] M. Fahim, V. Sharma, T.-V. Cao, B. Canberk, and T. Q. Duong, "Machine Learning-Based Digital Twin for Predictive Modeling in Wind Turbines," IEEE Access, vol. 10, pp. 14184–14194, 2022, doi: 10.1109/ACCESS.2022.3147602.

[22] Ildar Idrisov, "Digital Twin for State of Charge Estimation of a Vanadium Redox Flow Battery," in Paths to Sustainable Energy, InTech, 2010. doi: 10.5772/13338.

[23] F. Ibanez, A. Mahmoud, V. Yaroslav, V. Peric, and P. Vorobev, "Improving the power sharing transients in droop-controlled inverters with the introduction of an angle difference limiter," Int. J. Electr. Power Energy Syst., vol. 153, p. 109371, Nov. 2023, doi: 10.1016/j.ijepes.2023.109371.

[24] Kostyuchenko, L.P. "Imitatsionnoe modelirovanie sistemelektrosnabzheniya v programme MATLAB [Simulation modeling of power supplysystems in MATLAB]," Krasnoyar. gos. agrar. un-t [Krasnoyarsk State AgrarianUniversity], Krasnoyarsk, Krasnoyarsk State Agrarian University Publ., 2012, pp. 54-57. (in Russian).

[25] G. Biau and E. Scornet, "A Random Forest Guided Tour," Nov. 18, 2015, arXiv: arXiv:1511.05741. doi: 10.48550/arXiv.1511.05741.

[26] S. Hochreiter and J. Schmidhuber, "Long Short-term Memory," Neural Comput., vol. 9, pp. 1735–80, Dec. 1997, doi: 10.1162/neco.1997.9.8.1735.